\documentclass[useAMS,usenatbib]{mn2e}
\usepackage{epsfig}

\def\jupiter{{\rm J}}

\title[Does  GD\,356 have a Terrestrial Planetary Companion?]
{Does  GD\,356 have a Terrestrial Planetary Companion?}

\author[D. T. Wickramasinghe, J. Farihi, C. A. Tout, L. Ferrario \& R. J. Stancliffe]
{Dayal T. Wickramasinghe$^1$, Jay Farihi$^2$, Christopher A. Tout$^{1,3,4}$,
Lilia Ferrario$^1$
\newauthor
and Richard J. Stancliffe$^4$\\
$^1$Mathematical Sciences Institute, The Australian National University, ACT 0200, Australia\\
$^2$ Department of Physics and Astronomy, University of Leicester,
Leicester LE1 7RH\\
$^3$Institute of Astronomy, The Observatories, Madingley Road,
Cambridge CB3 0HA\\
$^4$Centre for Stellar and Planetary Astrophysics, Monash University,
PO Box 28M, Victoria 3800, Australia}
\begin{document}
\date{Accepted.  Received ; in original form} 
\pagerange{\pageref{firstpage}--\pageref{lastpage}} \pubyear{}

\maketitle

\label{firstpage}

\begin{abstract}

GD\,356 is unique among magnetic white dwarfs because it shows
Zeeman-split Balmer lines in pure emission.  The lines originate from
a region of nearly uniform field strength ($\delta B/B \approx 0.1$)
that covers $10\,$per cent of the stellar surface in which there is a
temperature inversion.  The energy source that heats the photosphere
remains a mystery but it is likely to be associated with the presence
of a companion.  Based on current models we use archival {\em Spitzer}
IRAC observations to place a new and stringent upper limit of
$12\,M_\jupiter$ for the mass of such a companion.  In the light of
this result and the recent discovery of a 115\,min photometric period
for GD\,356, we exclude previous models that invoke accretion and
revisit the unipolar inductor model that has been proposed for this
system.  In this model a highly conducting planet with a metallic core
orbits the magnetic white dwarf and, as it cuts through field lines, a
current is set flowing between the two bodies.  This current
dissipates in the photosphere of the white dwarf and causes a
temperature inversion.  Such a planet is unlikely to have survived the
RGB/AGB phases of evolution so we argue that it may have formed from
the circumstellar disc of a disrupted He or CO core during a rare
merger of two white dwarfs.  GD\,356 would then be a white dwarf
counterpart of the millisecond binary pulsar PSR\,1257+12 which is
known to host a planetary system.

\end{abstract}


\section {Introduction}

GD\,356 was first noted to be peculiar among magnetic white dwarfs
through the discovery of strong Zeeman split Balmer lines in pure
emission \citep{greenstein1985}. The number of known magnetic white
dwarfs has since increased to some $220$ through surveys such as
SDSS but GD\,356 remains unique in exhibiting these properties.

A detailed study of the atmosphere of GD\,356 by \citet{ferrario1997}
established that the line and continuum
spectra can be modelled with a white dwarf with an effective
temperature $T_{\rm eff} = 7\,500\,$K and an assumed gravity $\log_{10}
(g/{\rm cm\,s^{-2}}) = 8$ with a centred dipole field structure and
polar field strength $B_{\rm p} =13 \,$MG. The emission lines arise
from a ring like spherical sector or strip around the magnetic pole covering
$10\,$per cent of the stellar surface.  The lines are a result of a temperature
inversion that begins deep within the photosphere at optical depths of
$ 1.0$. The remainder of the photosphere produces absorption lines
broadened by the underlying dipolar field.  These lines are masked by
the emission lines in the flux spectra but are discernible in the
polarization spectra. Barring the possibility of a rapid rotator, the
observed lack of variability of the spectrum over periods of several
hours to days must indicate that the spin and dipole axes are nearly
aligned.  Likewise, the absence of reports of significant changes in
the occasional spectra taken over a $25\,$yr  period (G. Schmidt and J. Liebert, private communication) suggests that the
photospheric region that gives rise to the emission lines has a stable
structure on this time scale.

The fields of the high field magnetic white dwarfs (HFMWDs), of $1 <
B/{\rm MG} < 1000$, such as GD\,356 are generally believed to be of
fossilized origin rather than dynamo generated
\citep{wickramasinghe2000}.  They are most likely generated in a
common envelope that left a very close binary or merged core
\citep{tout2008}.  In this case all single white dwarfs with high
magnetic fields have evolved from binary stars that merged during
common envelope evolution or shortly afterwards.  Though very
uncertain, estimates of their number densities indicate that about
three times as many systems that enter a common envelope phase of
evolution end up merging as form cataclysmic variables.  The long
ohmic decay time scales, $8-12\,$Gyr \citep{cumming2002}, of the
dipolar component in magnetic white dwarfs and the lack of
observational evidence for differences in the mean dipolar field
strength along the white dwarf cooling sequence supports this
hypothesis.  Cool white dwarfs develop convective envelopes that could
potentially lead to generation of magnetic fields by a contemporary
dynamo but there is no evidence for an increase in the incidence of
magnetism among the cooler and more convective stars in the well
studied high-field group of white dwarfs.  It is of course possible
that all cool white dwarfs with outer convective envelopes have dynamo
generated fields that are below the current observational limit of
detectability from Zeeman polarimetry of $B\approx 10^3\,$G
\citep{jordan2007}.  However, attempts to detect X-ray emission from
coronae that may be generated through magnetic activity in such stars
have so far led to upper limits that are well below theoretical
predictions \citep{musielak2003}.

Since these early investigations there have been some new
observations of
this star.  \citet{brinkworth2004} reported the detection of low-amplitude
($\pm0.2 \,$per cent) near sinusoidal variability in the $V$ band with a
period of $115\,$min.  They attribute this to the rotation period of
the star. They presented a model in which a dark spot covers $10\,$per
cent of the stellar surface and is viewed nearly face on or edge on
as the star rotates.  They speculated that the temperature inversion
required to explain the spectroscopic data must occur in this
region. While the cause of the observed photometric variability may
well be related to a temperature differential between the line
emission region and the remainder of the star, the idea of a dark
magnetic spot presumably with an enhanced field caused by magnetic
activity as in the Sun and stable over some $25\,$yr is less
attractive. The modelling shows no evidence for an enhanced field
strength in the emission line region. Rather it is simply a specially
heated region of the star with an otherwise approximately dipolar
field structure.  It is thus more likely that GD\,356 has a fossil
field like other white dwarfs of similar field strength.

The source of energy that powers the emission line region in GD\,356
remains a mystery. The observed luminosity in the Balmer lines is
$2\times 10^ {27} \rm{ergs~s^{-1}}$, much larger than the stringent
upper limit of $6\times 10^{25} \rm{ergs~s^{-1}}$ that has recently
been placed on the X-ray luminosity of GD\,356.  So the
source of heating is not an X-ray corona \citep{weisskopf2007}.
Given the implausibility of a single star interpretation, the most
likely possibility is that the energy is extracted in some way from a
companion. The lack of evidence for an accretion disc in the line
spectrum or of emission from accretion shocks appears to preclude 
intermediate polar type and AM~Her type models.  

One of the more intriguing models that has been proposed for GD\,356
assumes that it has a companion of planetary mass with a conducting
composition in a close orbit \citep{li1998}.  Such a planet would act as
a unipolar inductor generating an electrical current that flows
between the two stars analogous to the model proposed for the
Jupiter-Io system \citep{goldreich1969}.  Ohmic dissipation in the
atmosphere of the white dwarf leads to a temperature inversion in a
ring surrounding the magnetic pole as the planet orbits the white
dwarf.  This model, as it presently stands, hinges rather heavily on
the possibility of the survival of such a planet through the RGB and
AGB phases of evolution.

In this paper we present an analysis of archival {\it Spitzer}/IRAC
observations of GD\,356 that rule out a companion of mass as greater
than $12\,M_\jupiter$, according to substellar cooling models. We then
discuss possible accretion models and the unipolar inductor model in
the light of the new mass limits and the recently discovered
photometric period for this star.  We argue that the ensemble of
presently available observations is inconsistent with accretion from a
gaseous planetary companion being the source of the anomalous line
emission seen in GD\,356.  We then show that rocky planets are
unlikely to survive the AGB phase of evolution and be dragged into a
close orbit as is required in the unipolar inductor model. We propose
that, for this model to be viable, the planet must have formed from
material in a disc that resulted from the merging of two white dwarfs
and argue that GD\,356 may be the result of such a rare event akin to
the event that resulted in the formation of planets around the
millisecond pulsar PSR\,1257+12 \citep{wolszczan1992}.

\begin{table}
\begin{center}
\caption{Mid-Infrared Fluxes for GD\,356\label{tbl1}} 
\begin{tabular}{@{}ccc@{}}
\hline
Wavelength		&Model Flux	&Measured Flux\\
$\lambda/\mu$m		&$F_\nu/\mu$Jy	&$F_\nu/\mu$Jy\\
\hline
3.6		&512	&$523\pm26$\\
4.5		&339	&$347\pm17$\\
5.7		&218	&$222\pm12$\\
7.9		&121	&$138\pm9$\\
\hline
\end{tabular}
\end{center}
There is a $2.0\sigma$ excess measured at $7.9\,\mu$m.

\end{table}

\section {The observations, limits on companion mass and implications}

\begin{figure}
\includegraphics[width=84mm]{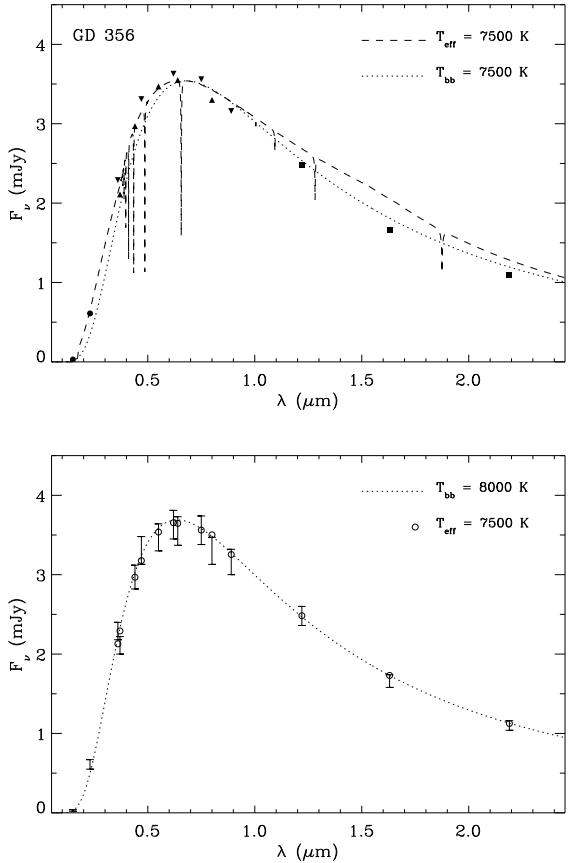}
\caption{Ultraviolet though near-infrared SED of GD\,356.  The upper
panel displays all available photometric data (filled symbols) for the
white dwarf plotted beside a 7\,500\,K hydrogen-rich model (dashed line) 
and a 7\,500\,K blackbody (dotted line).  Circles are {\em GALEX}, upward
triangles are $UBVRI$, downward triangles are $ugriz$ and squares are
$JHK$ photometry.  The lower panel shows fluxes for a $7\,500\,$K helium-rich 
model (open circles) at the optical and near-infrared bandpasses, together 
with an $8\,000\,$K blackbody model (dotted line).  This reproduces the helium 
atmosphere model fluxes rather well.  
\label{fig1}}
\end{figure}

\begin{figure}
\includegraphics[width=84mm]{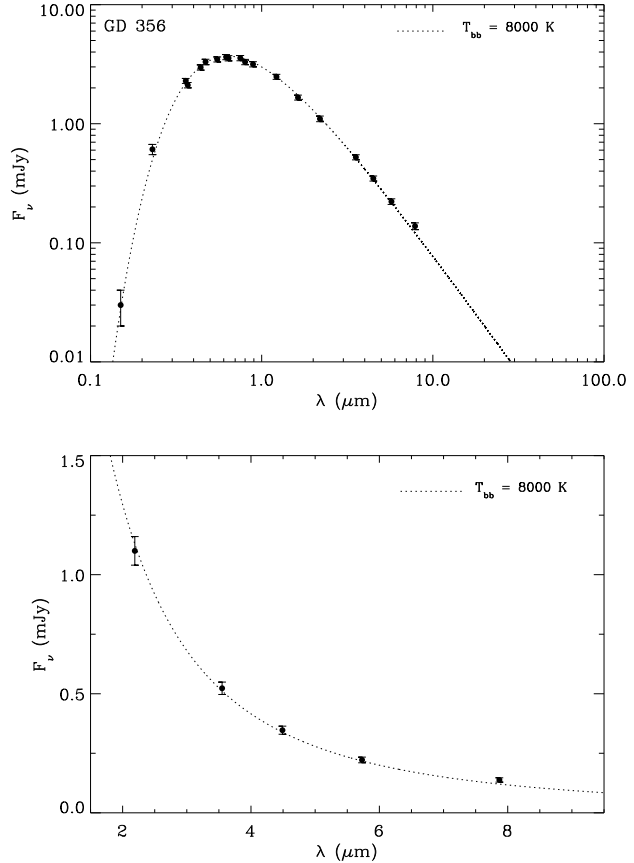}
\caption{Full SED of GD\,356.  The upper panel shows all photometry (filled circles), 
now including the {\em Spitzer} IRAC flux measurements and the $8\,000\,$K blackbody 
model, normalized to the same level as in Fig.~\ref{fig1}.  The lower panel is a
linear plot of the infrared data and the potential modest excess at 8\,$\mu$m.
\label{fig2}}
\end{figure}

We combine multi-wavelength photometry from several sources in order
to constrain the spectral energy distribution of GD\,356, particularly
the photospheric emission at infrared wavelengths.  Far- and
near-ultraviolet fluxes were obtained from the {\em Galaxy Evolution
  Explorer} \citep[{{\it GALEX}},][]{martin2005} data archive.  These
data are uncorrected for extinction and were assigned 10 to 30\,per
cent uncertainties (larger than their quoted errors) owing to this
fact.  Optical $BVRI$ photometry was taken from \citet{bergeron2001},
supplemented with $U$-band measurements in \citet{mccook2008}, while
$ugriz$ photometry were available from the Sloan Digital Sky Survey
Data Release~7 \citep[SDSS~DR7][]{abazajian2009}.  Near-infrared $JHK$
fluxes were taken from the weighted average of photometry from
\citet{bergeron2001} and the Two Micron All-Sky Survey
\citep[2MASS,][]{skrutskie2006}.  The optical and near-infrared fluxes
shown in Fig.~\ref{fig1} are all weighted equally with assumed
5\,per cent errors.

A thorough photometric and trigonometric parallax analysis of GD\,356
by \citet{bergeron2001} yields $T_{\rm eff}=7\,510$\,K, $\log_{10}(g/\rm
cm\,s^{-2}) = 8.14$ and a {\em helium}-rich atmosphere for the DAe
white dwarf.  The top panel of Fig.~\ref{fig1} shows our first attempt
to fit the data, represented by points, with a model photosphere with
both a 7\,500\,K, $\log_{10} (g/{\rm cm\,s^{-2}}) = 8.0$ DA spectral model \citep{koester2009} and
an identical temperature blackbody model.  Neither of these models is
sufficient to account for the entire spectral energy distribution
(SED) simultaneously.  In the absence of a full spectral model for helium
atmosphere white dwarfs cooler than $10\,000\,$K, in the lower panel
we plot the $UBVRI$ and $ugriz$ model fluxes as
open circles for a helium-rich white dwarf of 7\,500\,K and $\log_{10}(g/\rm
cm\,s^{-2}) = 8$ \citep{holberg2006,fontaine2001}, against the
photometry, represented as error bars.  On top of this is plotted an
8\,000\,K blackbody model which mimics the 7\,500\,K helium-rich model
fluxes quite well.  We use this blackbody fit to extrapolate towards
longer wavelengths and iterate with the measured mid-infrared fluxes.

Lastly we have analyzed archival {\em Spitzer} IRAC images of GD\,356
following the methods of \citet{farihi2009}.  It is worth remarking that
these images were free of potential flux-contaminating sources within
the $r=3\farcs6$ photometric aperture and that the signal-to-noise was
sufficiently high that the flux errors at all wavelengths are dominated
by IRAC calibration error.  These previously unpublished fluxes are
listed in Table~\ref{tbl1} and plotted in the upper panel of Fig.~\ref{fig2}
together with the shorter wavelength photometry and our
selected model.  The entire photometric SED, except possibly the $7.9\,\mu$m flux which
appears to be in excess at the $2.0\sigma$ level, is fitted well by the
$8\,000\,$K blackbody model.  The stellar image
at this longest wavelength is both highly symmetric and sufficiently
bright to show a faint Airy ring in the $0\farcs6$ pixel$^{-1}$ mosaic
but we are cautious about the interpretation of an excess at this
level without more data (see the lower panel of  Fig.~\ref{fig2}).
 
We follow \citet{farihi2008b} and use the measured photospheric flux
at $4.5\,\mu$m to place an upper limit to the mass of a possible
substellar companion to GD\,356.  For a white dwarf of mass
$0.67\,M_\odot$ and effective temperature $7\,500\,$K, the cooling age
is 1.6\,Gyr \citep{bergeron1995a}.  The main-sequence progenitor of
GD\,356 should have had a mass $M_{\rm MS} = 3.25\,M_\odot$ according
to the initial-to-final mass relation
\citep{williams2009,kalirai2008,dobbie2006} and hence an estimated
total lifetime of 2.1\,Gyr.  According to models, at the $d=21.1\,$pc
trigonometric parallax distance to the white dwarf, an unseen
$3\sigma$ (15\,per cent) excess at $4.5\,\mu$m places a companion
upper mass limit of $M_{\rm p} < 12\,M_\jupiter$ for an age of
2.1\,Gyr \citep[Baraffe 2007, private communication;][]{baraffe2003}.

We note there are no white dwarfs with {\em Spitzer}-only excesses owing
to companions \citep{farihi2009,farihi2008b,mullally2007}.  All known
infrared excesses from substellar companions reveal themselves in the
near-infrared by $2\,\mu$m at the latest \citep{steele2009,
farihi2008c,burleigh2006,farihi2005,becklin1988}.  These results give strict upper limits
to the mass of Jupiter-sized companions to white dwarfs of typically
between 5 and~$20\,M_\jupiter$ from young, less than $1\,$Gyr, to intermediate,
$2-5\,$Gyr, total ages.  The limits at GD\,356 are therefore
commensurate with other $4.5\,\mu$m excess searches.
	

The stringent upper limit that we have deduced for the mass of a
possible secondary places new constraints on a binary accretion model
for GD\,356.  First suppose that the substellar type companion is
gaseous with $M_{\rm p}\le 12\,M_\jupiter$.  We expect such a companion to be
tidally locked with the orbital period just as in cataclysmic
variables.  For a Roche lobe filling gaseous companion of mass $M_{\rm p}$
orbiting a white dwarf of mass $M_{\rm wd}$ with separation $a$, the
equivalent spherical Roche lobe radius $R_{L2}$ is given by
\begin{equation}
\frac{R_{L2}}{a}=\frac{2}{3^{\frac{4}{3}}}\left(\frac{M_{\rm p}}{M_{\rm p}+M_{\rm wd}}\right)^{\frac{1}{3}}
\label{roche}
\end{equation}
\citep{paczynski1971}.  For $M_{\rm p} \ll M_{\rm wd}$
Kepler's third law gives
\begin{equation}
R_{L2} = 0.05\left(\frac{M_{\rm p}}{10\,M_\jupiter}\right)^\frac{1}{3}
\left(\frac{P}{1\,\rm h}\right)^\frac{2}{3}\,\rm R_\odot.
\end{equation}
For $0.1 \le M_{\rm p}/M_\jupiter \le 70$, the radius of a planetary
secondary is $R_2 \approx  0.1\,R_\odot$
\citep{hubbard1994}.  If such a secondary is to just fill its Roche lobe or
lie within it the orbital period must satisfy
\begin{equation}
P\ge2.83\left(\frac{10\,M_\jupiter}{M_{\rm p}}\right)^\frac{1}{2}\,\rm h.
\end{equation}
The limit of $M_{\rm p} < 12\,M_\jupiter$ could therefore be
consistent with a gaseous planet in a binary system with an orbital
period $P_{\rm orb}\ge 2.7\,$h in which the white dwarf rotates at a
period of $115\,$min \citep{brinkworth2004} asynchronously with the
orbit.  Such a system could in principle be a post period bounce
cataclysmic variable in which the companion mass has been reduced to
low values by mass transfer.  In such a case the white dwarf need not
be tidally locked and should be rotating faster than the orbit because
it is still accreting angular momentum, albeit at a slow rate.
However, the time scale to reach post bounce periods of greater than
$2.7\,$h would typically exceed a Hubble time \citep{kolb1999} so that
this possibility can actually be eliminated given the youth of the
white dwarf.  Alternatively a gaseous planet may have been dragged in
to a close orbit during AGB evolution but not evaporated or it could
have formed in a disc following a merger of two stars.

However, for gaseous planets with $P_{\rm orb}\ge 2.7\,$h, the mass transfer rate
by Roche lobe overflow would be less than $10^{-13}\,M_\odot\,\rm{yr}^{-1}$
when the orbital evolution is driven by gravitational
radiation.  For this white dwarf the corresponding specific accretion
rate per unit area would be about $10^{-6}\,\rm g\,cm^{-2}\,s^{-1}$ if
accreting material were to flow on to the $10\,$per cent of the
surface area of the white dwarf where a temperature inversion is seen.
At these accretion rates a shock would not form and the atmosphere would
be heated by particle bombardment and cooled by cyclotron
emission.  Deep heating would not be expected in the bombardment
regime and heating to optical depths of order unity, as is required to
explain the emission lines in GD\,356, is excluded
\citep{ferrario1997}.  Accretion from a companion at higher rates,
perhaps by an irradiation induced wind, can also be rejected because
the usual
indicators of disc or funnel accretion are not seen in this
system either.

Eliminating these possibilities, we are once again led to favour the
hypothesis of a conducting planet orbiting the magnetic white dwarf,
with the excess emission powered by the unipolar inductor mechanism
\citep{li1998}.  The planet must have a conducting core and be free of
any atmosphere so that the inducted current dissipates in and heats
the white dwarf atmosphere.  At the mass of the Earth even a rocky
planet in a close orbit would assume the shape of a Roche potential
because tidal forces would be sufficient to melt any solid crust.  For
such a planet, of mean density of $\rho_{\rm p}$, the Roche limit
obtained with equation~(\ref{roche}) is
\begin{equation}
P^2 > 4\pi\left(3\over 2\right)^5{1\over G\rho_{\rm p}}.
\end{equation}
This allows orbital periods of greater than about $4.7\,$h
for $\rho_{\rm p} = 5\,\rm g\,cm^{-3}$.
We have estimated the contribution that this planet would make to the
observed energy distribution.  A rocky planet with an albedo
$\epsilon$ in orbit at a distance $a$ from the white dwarf, the
equilibrium effective temperature is given, to a first approximation, by
\begin{equation}
T_p =(1-\epsilon)^{1/4}\left(\frac{R_{\rm wd}}{2a}\right)^{1/2}
T_{\rm{wd}}
\end{equation}
For a bond albedo of $0.3$, we find a planet temperature of
$560\,$K  for an orbital period of $4$ hours.  
Such a planet would contribute $1\,$per cent of the white dwarf flux at 
$7.9\,\mu\rm m$ if it had a radius of $1.4 R_\oplus$. The $2\sigma$ 
excess observed at $7.9\,\mu\rm m$ could in
principle be consistent with the presence of the hypothesized rocky
planet, although other interpretations of this excess are also
possible.

\section{Discussion}

We must now ask how such a planet-like object can find itself so close to a
white dwarf.  We first show that it could not have been dragged in
from a larger orbit by the progenitor of the white dwarf and then
discuss the likelihood that it formed when a less massive companion
white dwarf merged with GD\,356.

\subsection {Planets orbiting white dwarfs}

The discoveries of Jovian planets orbiting evolved giant stars at
distances of the order of $1\,$au demonstrate that such planets can
survive at least the early giant phases of evolution.
\citet{lovis2007}
estimate that at least $3\,$per cent of evolved giant stars
($M\ge 1.8\,M_\odot$) have companions, including brown dwarfs, with
$M_{\rm p} \sin i > 5\,M_\jupiter$.  These results show that planets
form in intermediate stars that evolve into white dwarfs. Some
evidence that Jovian mass planets may survive through the RGB and AGB
phases to the white dwarf phase comes from the timing of pulsations in
ZZ~Cet stars.  It has been estimated that GD 66 may have a planet with
$M_{\rm p} \ge 2.11\,M_\jupiter $ orbiting at $2.4\,$au, although this
is based on only one measured turning point of the orbit and is therefore
not well constrained \citep{mullally2008}. Likewise V391~Pegasi, which
is an extreme horizontal branch star, appears to have a planet with
$M_{\rm p} \ge 3.2\,M_\jupiter$ orbiting at about $1.7\,$au
\citep{silvotti2007}.  However most SdB stars are in binaries so this
particular system is likely to have been the end product of binary
evolution.

Gaseous planets that are initially close enough to interact with the
expanding RGB or AGB star could simply be dragged in to a closer orbit
by bow-shock and tidal drag during an ensuing common envelope phase of
evolution or be completely destroyed by evaporation depending on their
initial mass.  The critical mass below which a Jovian planet is
expected to evaporate in the envelope of the giant star before the
envelope itself can be ejected is estimated to be about about
$15M_{\rm{Jup}}$ for a $1\,M_\odot$ star but there are large
uncertainties in this estimate related to parameters
such as efficiency of common envelope ejection
\citep{nelemans1998,siess1999}.  This estimate increases to
$120M_\jupiter$ for a $5\,M_\odot$ star.  
If the timescale for the common envelope phase were short enough it
might be envisaged that a larger mass planet might have been partially
evaporated down to about $10\,M_\jupiter$.  However, given that
evaporation should accelerate as the planet loses mass it is very
unlikely that such fine tuning could have occurred.
A main-sequence star--planet
system that evolves through these phases would be seen either as a
single white dwarf or as a white dwarf with a close planetary
companion with a mass above this critical value.  GD\,1400 and
WD\,0137 are close binaries with orbital periods of $10$ and~$2\,$h
respectively (\citep{maxted2006,farihi2004}).  GD\,1400 is a CO white dwarf, 
which is the remnant of
AGB phase of evolution, while WD\,1037 is a He white dwarf which has
only passed through the RGB phase of evolution.  The companions are
$50-60\,M_\jupiter$ brown dwarfs.  Thus the empirical evidence
appears to be that a companion of this mass can survive evaporation and
in-spiralling during RGB/AGB evolution, eject the envelope and be
seen as a close binary now. We note, however, that \citet{villaver2007} have questioned
whether such a close companion could survive the intense radiation to
which it would be subjected to by the newly formed white dwarf and have 
proposed that such systems are more likely to have arisen from a merger 
of two white dwarfs. Against this hypothesis is the observation that
main-sequence star--brown dwarf pairs appear to occur with about the
same frequency as white dwarf--brown dwarf pairs
indicating that the latter can form without the need for strong binary interaction.

The fate of rocky planets, particularly that of the Earth itself,
during the late stages of stellar evolution has been looked at in some
detail.  \citet{sackmann1993} thought that the Earth would escape
because mass loss increases the orbital separation faster than the Sun
grows on both the RGB and AGB.  \citet{rasio1996} went on to point out
that tides, if a little stronger than expected, induced by the Earth
on the Sun might actually cause the Earth to spiral in during the RGB
phase.  Like \citet{sackmann1993} they worked with Reimers'
\citep{reimers1975} mass-loss rates and so claimed that survival of
the RGB would lead to an orbit wide enough to survive the AGB too.
\citet{rybicki2001} used more realistic thermally pulsing AGB models
with weak mass loss on the RGB and so their stars experience many more
thermal pulses and grow fast enough to swallow the Earth before the
end of the AGB.  \citet{schroder2008} invoke even stronger mass loss
towards the luminous tip of the RGB so that their Sun grows large
enough to engulf a tidally spiralling Earth on the RGB but not the
AGB.  The orbital angular momentum of a low-mass terrestrial planet
that is engulfed during the RGB phase is insufficient to eject the
common envelope.  The orbit rapidly decays owing to tidal and
bow-shock drag and the planet plunges into the central star and is
destroyed.  A planet that is first engulfed only during a thermal
pulse on the AGB phase might have a better chance of survival.
\citet{rybicki2001} suggest that such a planet would be dragged in by
$10-70\,R_\odot$ per thermal pulse and is also likely to be destroyed.
\citet{willes2005} went back to the older AGB models of
\citet{sackmann1993} to argue that a fraction of such systems may
survive stellar evolution and end up as close companions to the white
dwarfs.  However these models had fewer thermal pulses and did not
grow as much on the AGB as is now thought.  Indeed \citet{willes2005}
relied on the star shrinking sufficiently over the last few pulses.
Our models \citep{stancliffe2004,stancliffe2008} show AGB stars
growing rapidly at the end of their lives after the onset of a
superwind.  In which case engulfment cannot be avoided if a planet is
dragged in.  While uncertainty remains in the evolution of AGB stars
it is possible that planets within a narrow range of separations from
their stars might actually be dragged in to closer orbits.  However
the fine tuning of the orbital decay that would be required from pulse
to pulse makes it very unlikely that the orbit could be reduced as
much as required for GD\,356.

Assuming that any planet that might be sufficiently dragged in would
be engulfed by the star's envelope, we must then consider the
implicit assumption that an Earth-like planet is likely to evaporate
during such periods of engulfment and so would not survive.
To totally evaporate a planet, of mass $M_{\rm p}$ and radius $R_{\rm
p}$, it must absorb sufficient thermal energy to overcome its
gravitational binding energy
\begin{equation}
E_{\rm gr} = {\zeta GM_{\rm p}^2\over R_{\rm p}} =
10^{38}\zeta\left(M_{\rm p}\over M_\oplus\right)^2\left(R_\oplus\over
R_{\rm p}\right)\,\rm erg,
\end{equation}
where $\zeta = 3/5$ for a uniform density body.  For a typical rocky
planet this is much larger than the total energy, of less than
$10^{36}\,$erg, to sublimate, dissociate and ionize the planet.  As
long as the planet's core remains at a temperature significantly lower
than the ambient temperature bath the timescale for evaporation is
just the Kelvin-Helmholtz timescale for the planet if it were
radiating as a black body with the temperature of the bath.  Thus the
evaporation timescale
\begin{equation}
t_{\rm evap} = {\zeta GM_{\rm p}^2/R_{\rm p}\over 4\pi\sigma R_{\rm p}^3T_*^4},
\label{evaptime}
\end{equation}
where $T_*$ is the temperature of the ambient giant stellar material.
In equilibrium the thermal energy of the planet would exceed the
gravitational binding energy when
\begin{equation}
{3\over 2}{R\over\mu}T_{\rm eq} M_{\rm p}> E_{\rm gr},
\end{equation}
where $R$ is the gas constant.  The mean molecular weight $\mu$
depends on the composition and ionization state of the planetary
material.  At $10^4\,$K most constituents of the Earth are at least
singly ionized and for a typical composition this gives $\mu\approx
15$.  At higher temperatures $\mu$ approaches two.  Thus
\begin{equation}
T_{\rm eq} < 30\,000 {M_{\rm p}\over M_\oplus}{R_\oplus\over R_{\rm p}}K.
\end{equation}
Thus for an Earth-like planet at a depth that takes it to $T_* >
30\,000\,$K we may apply equation~(\ref{evaptime}) and
$t_{\rm evap}$ is only about two days so the planet could not survive
such conditions.  

\begin{figure}
\epsfig{file=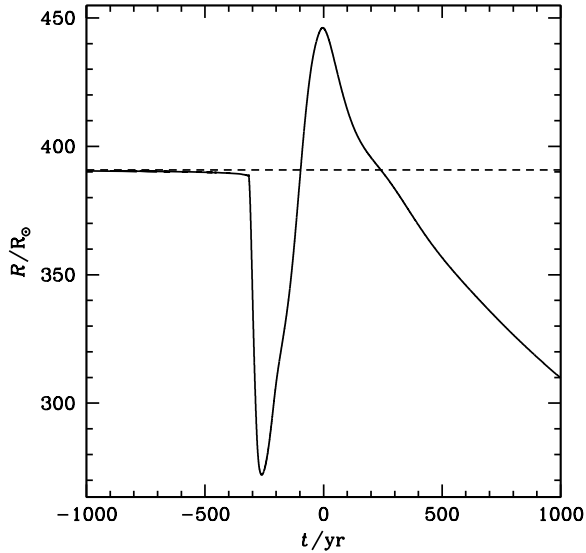,width=7.8cm}
\caption{\label{radage}
The evolution of radius with time during the fifteenth
thermal pulse of an initially $3\,M_\odot$ star when it has a core
mass of $0.65\,M_\odot$.  The star is beginning to lose mass in a
superwind and so is expanding rapidly from pulse to pulse.  The dashed
line marks the maximum radius reached during the previous interpulse
period.  A planet at this radius is engulfed for more than $300\,$yr
and reaches a depth of more than $50\,R_\odot$ in the AGB envelope.}
\end{figure}

\begin{figure}
\epsfig{file=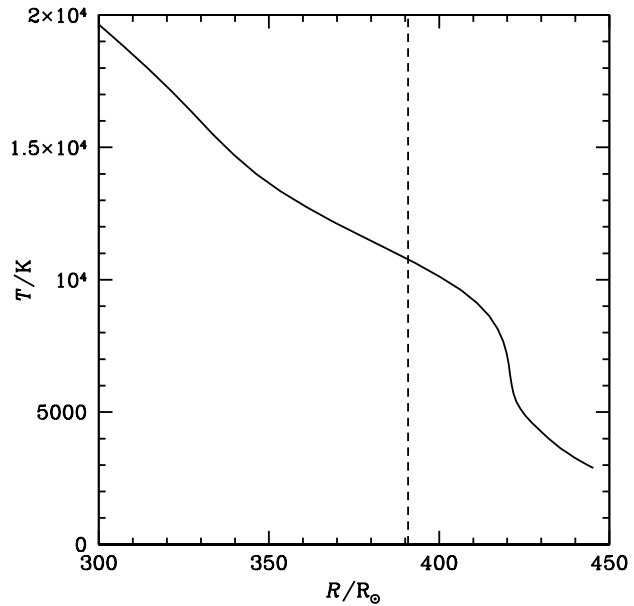,width=8.4cm}
\caption{\label{temprad} 
The temperature structure with radius for the fifteenth pulse of our
initially $3\,M_\odot$ at $t = 0$ in Fig.~\ref{radage} when the star
reaches its maximum radius.  The dashed line marks the maximum radius
reached during the previous interpulse period and so corresponds to
the depth of a planet that just survived the fourteenth pulse.}
\end{figure}

Fig.~\ref{radage} illustrates the evolution of the stellar radius of
an initially $3\,M_\odot$ star with a current core mass of
$0.65\,M_\odot$ through a single AGB pulse calculated with the
Cambridge Stars code \citep{stancliffe2004}.  A planet that is located
just outside the star prior to a pulse is engulfed by the expanding
envelope for a period of more than $300\,$yr.  Even if the planet were
not dragged in, Fig.~\ref{temprad} demonstrates that it would be
exposed to ambient temperatures of $10{,}000\,$K for long enough to
evaporate the more volatile elements.  As the pulses proceed the mass
of the planet would fall and the temperature and depth in the stellar
envelope required for complete evaporation would rapidly diminish.  We
would not expect a rocky Earth-like planet to survive more than the
first few pulses that engulf it.  If the planet is dragged in it
is exposed to yet higher temperatures for even longer.  The
temperature at the base of an AGB star's envelope is in excess of
$10^7\,$K so a planet dragged into a close orbit within the giant
envelope could not survive.

For the extreme case of a planet engulfed at the very end of the
thermally pulsing AGB when only about $0.05\,M_\odot$ of stellar
envelope remains the planet might survive partial immersion.  However
the binding energy of such an envelope is only about one thousandth of
that required to bring the planet in to a close enough orbit around
the white dwarf.  We therefore conclude that a rocky, Earth-like planet
cannot both survive evaporation and end up in a very close orbit.

\subsection{Merging double white dwarfs}

We consider a binary evolution scenario which leads to the formation
of a CO white dwarf with a lower mass He or CO white dwarf companion
through common envelope evolution.  The two stars are subsequently
drawn together by gravitational radiation and merge.  During the
merging this companion breaks up and forms a massive disc around the
remaining CO white dwarf.  Such a disc could be composed of CO-rich or
He-rich material.  The mass of the disc is somewhat more than one
tenth of that of the accreting star and so becomes unstable to its own
self gravity.  The disc expands and cools as the central star accretes
matter.  When the temperature in the outer disc is cool enough dust
and rocks form and these clump to form a rocky planetary core. This
model is very similar to that initially proposed to explain planetary
companions to millisecond pulsars \citet{podsiadlowski1991}.

In order for the two white dwarfs to merge the common envelope process
must leave them close enough for gravitational radiation to act
quickly.  \citet{tout2008} have demonstrated how high magnetic fields
in white dwarfs are almost certainly generated in common envelopes
from which the cores emerge already close together.  Thus the high
field in this system is evidence that the white dwarf most likely
emerged from common envelope evolution with a close companion, in this
case a second, less massive white dwarf.  The mass of GD\,356 of
$0.67\,M_\odot$ is already above the average for CO white dwarfs.  In
order to leave two white dwarfs that can merge the system must
originally have been close enough that the evolution of both stars
was curtailed by mass transfer.  Thus we might envisage mass transfer
from the initially more massive star to begin when it has a CO core of
say $0.4\,M_\odot$.  If this were to lead to a mild common envelope
phase the orbit would then shrink so that the second star fills its
Roche lobe early on red giant branch with a He core of about
$0.3\,M_\odot$.  Alternatively the first star might have filled its
Roche lobe as a subgiant, evolved through an Algol phase to a helium
white dwarf.  Its rejuvenated companion could then go on to fill its own
Roche lobe on the AGB followed by common envelop evolution that leaves
its CO core in a close orbit.  In either case for the final common
envelope must leave the two cores sufficiently close that
they can be driven together by gravitational radiation and their mass
ratio must be less than 0.628 so that the ensuing mass transfer is
dynamically unstable.

When two white dwarfs merge, the more massive, being larger in radius,
fills it's Roche lobe first.  If the masses are sufficiently different
stable mass transfer could follow with the orbit widening as the
mass-losing star grows in radius.  However if the donor is more than
$0.628$ times the mass of the accretor the mass transfer is unstable
because the white dwarf grows faster than its Roche lobe expands.  In
this case numerical simulations show that the less massive white dwarf
is indeed tidally disrupted and accreted on to the more massive in
through a thick accretion disk
\citep{mochkovitch1989,benz1990,guerrero2004}.  So that the natural
outcome is a hot white dwarf surrounded by a thin remnant disc that
contains most of the angular momentum.  The nature of this disc is
likely to be very unusual, given that white dwarfs are composed
primarily of carbon and oxygen, rather than the hydrogen and helium of
more traditional circumstellar discs.

The formation of planets in such discs around neutron stars and white
dwarfs has been discussed by \citet{hansen2002} and \citet{livio2005}.
The expected outcome depends rather critically on the viscosity of the
disc.  Initially it has an outer radius of $10^9 - 10^{11}\,$cm,
determined by the orbital angular momentum of the disrupted companion.
As accretion proceeds the disc expands and cools.  While the viscosity
is determined by the magneto-rotational instability it is strong only
when the disc is ionized and the viscosity is negligible outside this
region.  If sufficient gas persists \citet{hansen2002} then finds that
planets form in the quiescent outer disc but at a higher temperature
than for a hydrogen rich composition, because of the higher ionization
potential of carbon and oxygen, by processes similar to those that are
believed to have occurred in the early solar system
\citep{lissauer1993}.  They predict that planets of $30 -300M_\oplus$
form in CO rich discs and are located within $0.2\,$au.  A similar
scenario is appropriate for He rich discs.  Planet formation may take
$10^8\,$yr.  The temperature of the white dwarf indicates that it has
been cooling in excess of $10^9\,$yr which easily accommodates this
along with sufficient time for any remnant disc to disperse.  We might
further speculate that all volatiles, and perhaps some transitional
elements, such as magnesium and silicon, may evaporate and be lost in
the very near environment of a hot and relatively luminous, newly
merged WD, while refractories, such as calcium, titanium and
aluminium, would more likely be retained, perhaps leading to a
refractory-metal planet.  In particular much of the oxygen that would
otherwise form oxides would almost certainly evaporate and be blown
away by radiation pressure so that metals that would otherwise oxidize
would be left to form a substantial metallic core.  Such a planet
might be more like Mercury than the Earth in composition.

Such a planet would have a metallic core of mass $0.03-1M_\oplus$ (for
a CO rich composition).  Once within $10\,R_\odot$ of the highly
magnetic white dwarf orbital energy can drive the unipolar induction
current \citep{li1998}.  Owing to its atmosphere the planet's
effective resistivity is initially larger than that of the white dwarf
so that energy, which is extracted from the orbit, is mainly
dissipated in the planet during the early phases of the magnetic
interaction.  As the planet drifts in, this heating facilitates the
evaporation of its atmosphere, until the effective resistivity of the
planet becomes smaller than that of the white dwarf's atmosphere.  The
heating then occurs mainly in the white dwarf atmosphere and the model
presented by \citet{li1998} for GD\,356 becomes applicable.

Some observational support for the possibility of the formation of a
second generation of planets around a star that is evolving into a
white dwarf has been provided by the discovery of a substantial
classical T~Tauri type dust disc surrounding an accreting first giant
ascent giant star TYC\,4144-329-2 \citep{melis2009}.  It has been
speculated that the observed disc may have resulted from the common
envelope interaction with a low mass stellar or substellar companion.

\section{Conclusions}

We have presented an analysis of archival {\it Spitzer} observations
of GD\,356 that shows that a firm upper limit of $12M_\jupiter$ can be
placed on the mass of a possible companion. The new observations
places further constraints on the orbital parameters of a possible
binary companion. In agreement with previous investigators, we have
again argued that accretion heating due to mass transfer from a
companions is an implausible mechanism for explaining the anomalous
line emission in this star. The unipolar inductor model, which
requires GD\,356 to have a rocky planet with a metallic core as a
companion, remains the best explanation for the peculiar properties of
this unique star.

Theoretical estimates indicate that it is unlikely that an Earth type
planet that was orbiting the main-sequence progenitor of a white dwarf,
at a distance that would allow it to be engulfed by the expanding
envelope during RGB/AGB phases of evolution, would survive the
subsequent evolution of the parent star and be seen as a close
companion to the white dwarf.  Such a planet would either be
evaporated as it is dragged into the core of the star during the
RGB/AGB phases of evolution or be left in an orbit at a much larger
radius of several hundred solar radii.  If the unipolar inductor model
for GD\,356 is to be viable an alternative origin must be sort for
its close companion. We have argued that the planet probably formed from
in an accretion disc following the disruption of a  white dwarf 
in a rare merger event. GD\,356 would thus be the white dwarf counterpart
of the millisecond binary pulsar PSR1257+12 which is known to
host a planetary system.
 
In conclusion, we note that the unipolar inductor model for GD 356
makes a definite prediction that could be verified by future
observations. Two fundamental and distinct periods, the rotation
period of the white dwarf and the orbital period of the planet around
the white dwarf, are expected to be seen in the emission line flux or
in any component of the continuum flux attributable to the heating.
\citet{livio1992} argue that planets are most likely to be found
around massive white dwarfs because these are more likely to have
merged in the past.  We would add that the white dwarf should not only
be massive but also posses a high magnetic field, created during the
common envelope evolution that must have preceded the merging.  It is
around such white dwarfs that the search for planets should be
concentrated and, if metallic, such systems might also show up as unipolar
inductors.

\section*{Acknowledgements}

CAT thanks Churchill College for his fellowship and Profs Dayal
Wickramasinghe and John Lattanzio for invitations to work in
Australia.  We thank Herbert Lau for critical comments on the survival
of rocky planets.

This publication makes use of data products from the Two Micron All
Sky Survey, which is a joint project of the University of
Massachusetts and the Infrared Processing and Analysis
Centre/California Institute of Technology, funded by NASA and the
National Science Foundation.  This work includes data taken with the
NASA Galaxy Evolution Explorer, operated for NASA by the California
Institute of Technology under NASA contract NAS5-98034.  Some data
presented herein are part of the Sloan Digital Sky Survey, which is
managed by the Astrophysical Research Consortium for the Participating
Institutions (http://www.sdss.org/).

\label{lastpage}
\end{document}